\documentclass{ws-procs9x6}

\title{Double transverse-spin asymmetries in Drell--Yan and $J/\psi$ production
from proton--antiproton collisions}

\author{
  M.~Guzzi$^{1,2,3}$,
  V.~Barone$^{4,5}$,
  A.~Cafarella$^{6}$, \\
  C.~Corian\`{o}$^{1,2}$
  \lowercase{and} P.G.~Ratcliffe$^{3,7}$
}

\address{%
  $^1$Dip.\ di Fisica, Universit\`{a} di Lecce, 73100 Lecce, Italy \\
  $^2$INFN, Sezione di Lecce, 73100 Lecce, Italy \\
  $^3$Dip.\ di Fisica e Matematica, Universit\`{a} dell'Insubria, 22100 Como, Italy
  \\
  $^4$Di.S.T.A., Universit\`{a} del Piemonte Orientale ``A.~Avogadro'' \\
      15100 Alessandria, Italy \\
  $^5$INFN, Gruppo Collegato di Alessandria, 15100 Alessandria, Italy \\
  $^6$Dept.\ of Physics, University of Crete, 71003 Heraklion, Greece \\
  $^7$INFN, Sezione di Milano, 20133 Milano, Italy
}

\begin{document}

\maketitle

\abstracts{We perform a NLO numerical study of the double transverse-spin
asymmetries in the $J/\psi$ resonance region for proton--antiproton collisions.
We analyze the large $x$ kinematic region, relevant for the proposed PAX
experiment at GSI, and discuss the implication of the results for the extraction
of the transversity densities.}

\section{Introduction}

The purpose of this talk is to illustrate a numerical analysis of the double
transverse-spin asymmetries in Drell--Yan processes in the $J/\psi$ resonance
region and to discuss the results with regard to the proposed PAX experiment and
the possibility of accessing the transversity densities in proton--antiproton
collisions.

\section{Access to transversity densities}

The missing leading-twist piece in the QCD perturbative description of the
nucleon is the transversity density,\cite{BDR} which is defined as the difference of
probabilities for finding a parton of flavour $q$ at energy scale $Q^2$ and
light-cone momentum fraction $x$ with its spin aligned $(\uparrow \uparrow)$ or
anti-aligned $(\uparrow \downarrow)$ with that of transversely polarized parent
nucleon
\begin{equation}
\Delta_T q(x,Q^2)=q_{\uparrow \uparrow}(x,Q^2)-q_{\uparrow\downarrow}(x,Q^2)\,.
\end{equation}
Given its chirally odd nature, transversity may be accessed in collisions
of two transversely polarized nucleons (Drell--Yan) via the double
transverse-spin asymmetries which are defined as the ratio
\begin{equation}
  A_{TT} =
  \frac{\sigma_{\uparrow \uparrow}-\sigma_{\uparrow \downarrow}}
       {\sigma_{\uparrow \uparrow}+\sigma_{\uparrow \downarrow}}
  = \frac{\Delta_T\sigma}{\sigma_{\rm unp}}
\end{equation}
between the transversely polarized and unpolarized cross-sections.

\begin{figure}\relax
  \centerline{\epsfxsize=3.5in\epsfbox{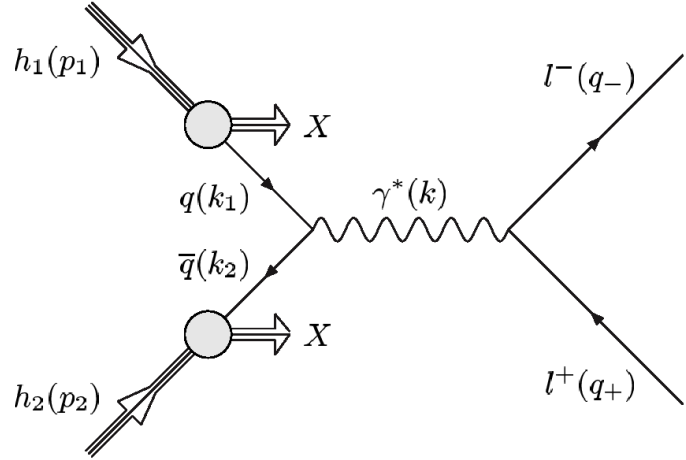}}
  \caption{Drell--Yan process
  \label{DY1}}
\end{figure}

Doubly polarized Drell--Yan production (illustrated in Fig.~\ref{DY1}) is the
cleanest process for probing transversity distributions. It has recently been
suggested that collisions of transversely polarized protons and antiprotons
should provide a very good opportunity to determine the nucleon transversity via
measurement of $A_{TT}$.

Double transverse-spin asymmetries depend may only on quark and antiquark
transversity distributions
\begin{equation}
  A_{TT} =
  \frac{\sigma_{\uparrow \uparrow}-\sigma_{\uparrow \downarrow}}
       {\sigma_{\uparrow \uparrow}+\sigma_{\uparrow \downarrow}}
  =
  \hat{a}_{TT}(\varphi)
  \frac{\sum_{q}e_q^2 \, \Delta_T q(x_1,M^2) \, \Delta_T\bar{q}(x_2,M^2) +
 (1 \leftrightarrow 2) }
       {\sum_{q}e_q^2 \, q(x_1,M^2) \, \bar{q}(x_2,M^2) +(1 \leftrightarrow 2) }
  \,,
\end{equation}
where $\hat{a}_{TT}(\varphi)$ contains the azimuthal angular dependence
\begin{equation}
  \hat{a}_{TT}(\varphi)=\tfrac{1}{2}\cos{2\varphi}
\end{equation}
and $M$ is the dilepton invariant mass.

Measurement of $A_{TT}^{pp}$ in the case of proton--proton collisions is planned
at RHIC but the asymmetry is expected to be small
(2--3\%).\cite{Barone1,Martin1} In fact, $A_{TT}^{pp}$ contains antiquark
distributions and the RHIC kinematics ($\sqrt{s}=200$\,GeV, $M<10$\,GeV, $x_1
x_2=M^{2}/s<3\times 10^{-3}$) probes the low-$x$ region where, compared to
$q(x)$, $\Delta_T{q}(x)$ is suppressed by QCD evolution. Such problems may be
avoided by measuring $A_{TT}^{p\bar{p}}$ in proton--antiproton collisions at
lower centre-of-mass energies;\cite{Barone1,Anselmino1} this is the program of
the PAX experiment at GSI.\cite{GSI1} For the GSI kinematics we have $s=30$ or
$45$\,GeV$^{2}$ in fixed-target, and $s=200$\,GeV$^{2}$ in collider mode,
$M>2$\,GeV and $\tau=x_1x_2=M^2/s>0.1$.

The GSI kinematics is such that the asymmetries for double transverse Drell--Yan
proton--antiproton processes are dominated by valence distributions and thus
probe the product $\Delta_T{q}\times\Delta_T{q}$. At LO we can write
\begin{equation}
  A_{TT}^{p\bar{p}}=\hat{a}_{TT}
  \frac{\sum_{q}\!e_q^2 [\Delta_T q(x_1,M^2)\Delta_T q(x_2,M^2)+
                         \Delta_T\bar{q}(x_1,M^2)\Delta_T\bar{q}(x_2,M^2)]}
  {\sum_{q}\!e_q^2 [q(x_1,M^2)q(x_2,M^2)+\bar{q}(x_1,M^2)\bar{q}(x_2,M^2)]}
  \label{LOAtt}
\end{equation}
and $A_{TT}^{p\bar{p}}/\hat{a}_{TT}$ is found to be of order of
$30\%$.\cite{Anselmino1,Efremov1}

\begin{figure}\relax
  \centering
  \resizebox*{85mm}{!}{\rotatebox{-90}
  {\includegraphics{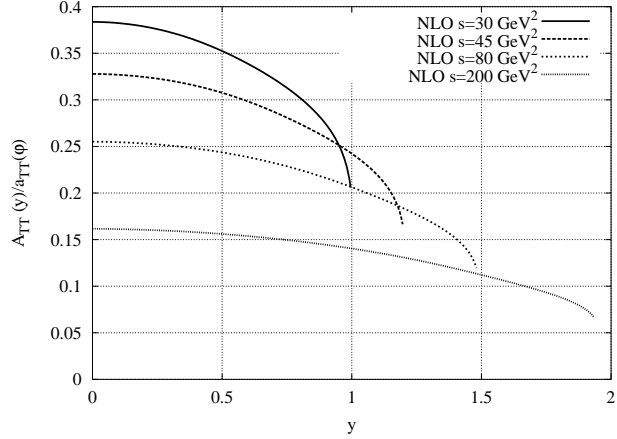}}}
  \caption{$A_{TT}(y)/\hat{a}_{TT}(\varphi)$ at NLO, with $M$ integrated
  from $2$ to $3$\,GeV using GRV input with the minimal bound
  $\Delta_T{q}(x,\mu)=\Delta{q}(x,\mu)$.}
  \label{asy_sc1}
\end{figure}

\begin{figure}\relax
  \centering
  \resizebox*{85mm}{!}{\rotatebox{-90}
  {\includegraphics{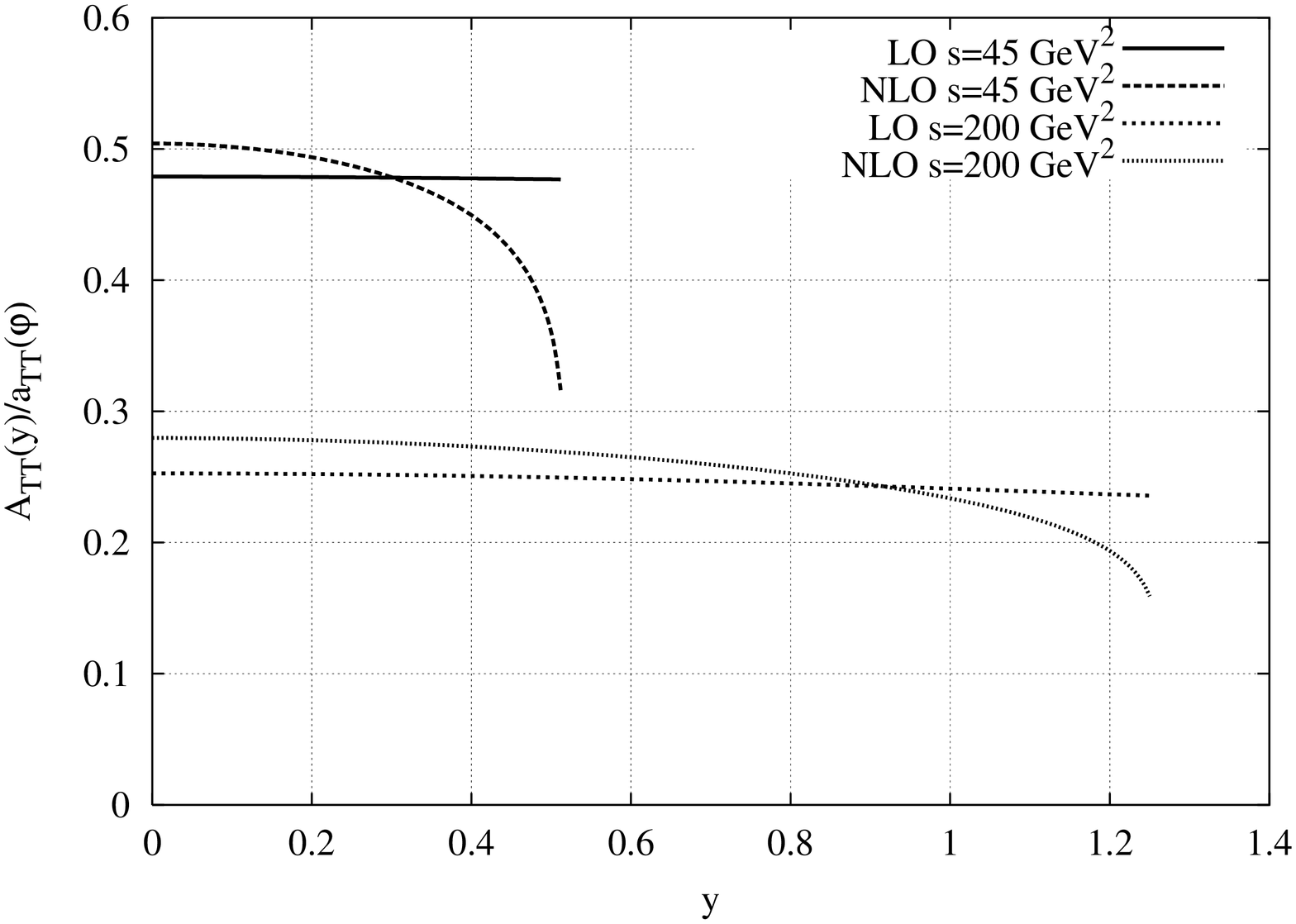}}}
  \caption{$A_{TT}(y)/\hat{a}_{TT}(\varphi)$ at LO (solid curve)
  \emph{vs.} NLO (dashed curve) at $M=4$\,GeV and $s=45$\,GeV$^2$ and
  $A_{TT}(y)/\hat{a}_{TT}(\varphi)$ at LO (dotted curve) \emph{vs.} NLO
  (dot-dashed curve) at $M=4$\,GeV and $s=200$\,GeV$^2$ using GRV input with
  the minimal bound.}
  \label{asy_sc4}
\end{figure}

At NLO the factorization formula of the cross-section for dilepton production in
transversely polarized proton--antiproton scattering
is\,\cite{Martin2,Mukherjee1}
\begin{multline}
  \frac{\text{d}\Delta_T\sigma}{\text{d}M \, \text{d}y \,\text{d}\varphi}
  =
  \sum_q e_q^2 \int_{\xi_1}^1 \text{d}x_1 \int_{\xi_2}^1 \text{d}x_2
  \left[ \Delta_T q(x_1,\mu^2) \Delta_T q(x_2,\mu^2) \right.
\\
  + \left. \Delta_T\bar{q}(x_1,\mu^2) \Delta_T\bar{q}(x_2,\mu^2) \right]
  \frac{\text{d}\Delta_T\hat\sigma}{\text{d}M \, \text{d}y \, \text{d}\varphi}
  \,,
  \label{fact}
\end{multline}
where $\mu^2$ is the factorization scale, $y$ is the rapidity of the dilepton
pair and the momentum fractions $\xi_1$ and $\xi_2$ are defined as
\begin{equation}
  \xi_1 = \sqrt{\tau} \, e^y, \qquad
  \xi_2 = \sqrt{\tau} \, e^{-y}, \qquad
  y = \frac{1}{2} \, \ln \frac{\xi_1}{\xi_2}\,.
  \label{kin}
\end{equation}
The NLO hard-scattering cross-section is
\begin{align}
  \frac{
    \text{d}\Delta_T\hat\sigma^{(1),\overline{\text{MS}}}
  }{
    \text{d}M \, \text{d}y \, \text{d}\varphi
  }
  \hspace{-5em} & \hspace{5em} =
  \frac{2\alpha^2}{9 s M} \, C_F \,
  \frac{\alpha_s(\mu^2)}{2\pi} \,
  \frac{4\tau(x_1x_2+\tau)}{x_1x_2(x_1+\xi_1)(x_2+\xi_2)} \,
  \cos(2\varphi)
  \nonumber
\\
  & \null \times
  \left\{
  \delta(x_1-\xi_1)\delta(x_2-\xi_2) \left[
  \frac{1}{4}\ln^2\frac{(1-\xi_1)(1-\xi_2)}{\tau}
  +\frac{\pi^2}{4}-2 \right]
  \right.
  \nonumber
\\
  & \hspace{1em} \null +
  \delta(x_1-\xi_1) \bigg[ \frac{1}{(x_2-\xi_2)_{+}\!\!}
  \ln\!\frac{2x_2(1-\xi_1)}{\tau(x_2+\xi_2)}
  + \!\left(\frac{\ln(x_2-\xi_2)}{x_2-\xi_2}\right)_{\!\!+}\!\!
  + \frac{\ln(\xi_2/x_2)}{x_2-\xi_2\!}\bigg]
  \nonumber
\\
  & \hspace{1em} \null +
  \frac{1}{2[(x_1-\xi_1)(x_2-\xi_2)]_{+}}
  + \frac{(x_1+\xi_1)(x_2+\xi_2)}{(x_1 \xi_2+x_2\xi_1)^2}
  -
  \frac{3\ln\left(\frac{x_1x_2+\tau}{x_1\xi_2+x_2\xi_1} \right)}
       {(x_1-\xi_1)(x_2-\xi_2)}
  \nonumber
\\
  & \hspace{1em} \null + \left.
  \ln \frac{M^2}{\mu^2} \left[
  \delta(x_1-\xi_1)\delta(x_2-\xi_2)
  \left( \frac{3}{4}+\frac{1}{2}\ln \frac{(1-\xi_1)(1-\xi_2)}{\tau} \right)
  \right. \right.
  \nonumber
\\
  & \hspace{12em} \null + \left. \left.
  \delta(x_1-\xi_1)\frac{1}{(x_2-\xi_2)_{+}} \right] \right\}
  + \Big[ 1 \longleftrightarrow 2 \Big]
  .
\label{sigmanlo}
\end{align}
In order to predict asymmetries, some assumption for the transversity
distributions is needed. For instance, we may take transversity equal to
helicity at some low scale (as suggested by certain models)
\begin{equation}
  \Delta_T f(x,\mu) = \Delta f(x,\mu) \quad \text{(minimal bound)}
\end{equation}
or, alternatively, saturation of the Soffer inequality\,\cite{Soffer}
\begin{equation}
  2 \, \vert \Delta_{T} f(x,\mu) \vert
=  f(x,\mu) + \Delta f(x,\mu)
  \,.
\end{equation}
We use NLO GRV input densities,\cite{GRV} with starting scale $\mu=0.63$\,GeV.
The relation between transversity and the GRV distributions is set at this
scale. QCD evolution is performed via the appropriate NLO DGLAP
equations.\cite{Cafacor1,Cafacorguz} In Fig.~\ref{asy_sc1} we see that in the
energy range relevant for the PAX experiment the asymmetries are around $35\%$.
\begin{figure}\relax
  \centering
  \resizebox*{85mm}{!}{\rotatebox{-90}
  {\includegraphics{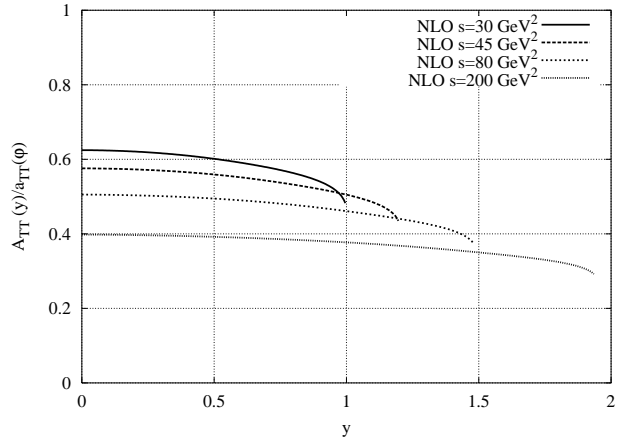}}}
  \caption{$A_{TT}(y)/\hat{a}_{TT}(\varphi)$ at NLO, with $M$ integrated
  from $2$ to $3$\,GeV using GRV input and saturating the Soffer bound.}
  \label{asy_sc2}
\end{figure}
From Fig.~\ref{asy_sc2} we see that in the case of the Soffer bound the
asymmetries are systematically larger than the asymmetries obtained in the case
of the minimal bound.
\begin{figure}\relax
  \centering
  \resizebox*{85mm}{!}{\rotatebox{-90}
  {\includegraphics{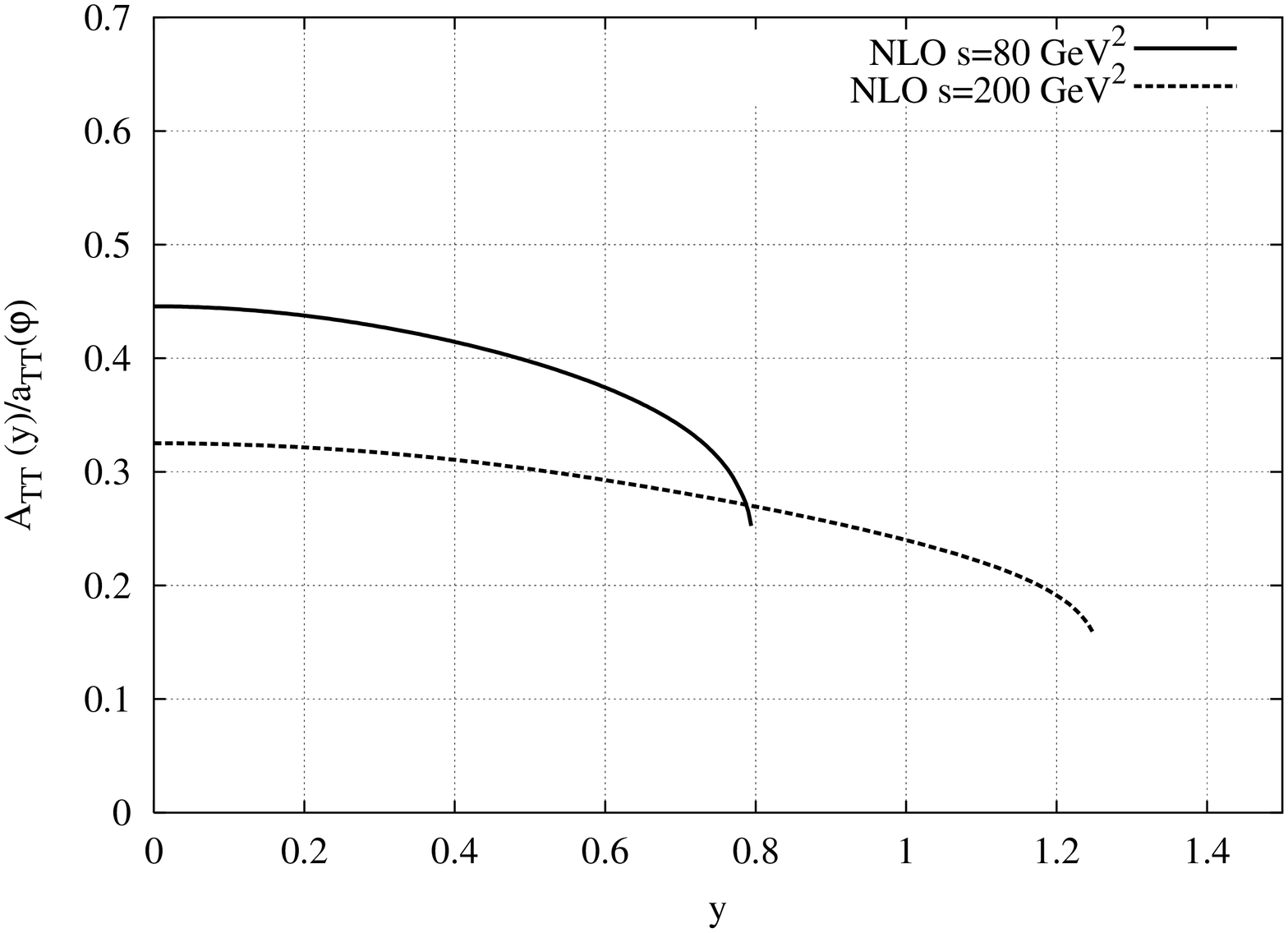}}}
  \caption{$A_{TT}(y)/\hat{a}_{TT}(\varphi)$ at NLO with $M$ integrated from $4$
  to $7$\,GeV using GRV input with the minimal bound.}
  \label{asy_sc3}
\end{figure}
In Fig.~\ref{asy_sc3} we display the asymmetry at larger $M$, where it grows up
to 45\% (but one should recall that the cross-section falls rapidly as $M$
increases). Applying the constraint $\Delta_T f(x,\mu)=\Delta f(x,\mu)$ at, say,
$1$\,GeV instead of $0.63$\,GeV would produce slightly larger asymmetries; this
is due to QCD evolution effects since $\Delta_Tf(x,\mu)$ is less suppressed by
evolving from $1$\,GeV than from $0.63$\,GeV. The comparison between LO and NLO
results is shown in Fig.~\ref{asy_sc4}, where one sees that NLO corrections have
very little affect on the asymmetries.
\begin{figure}\relax
  \centering
  \resizebox*{85mm}{!}{\rotatebox{-90}
  {\includegraphics{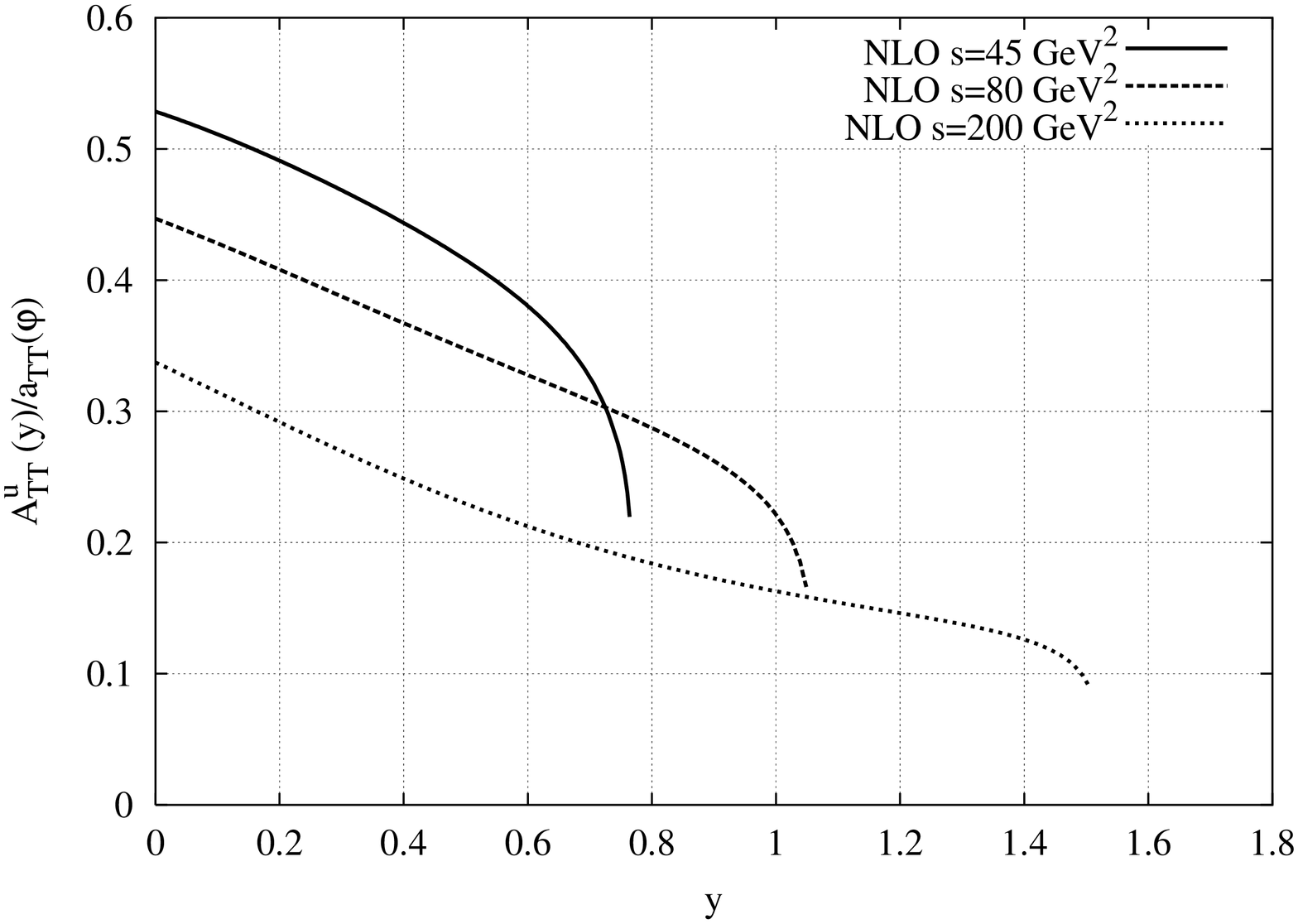}}}
  \caption{The double transverse-spin asymmetry in the $J/\psi$ resonance region
  for various c.m.\ energies. As usual, the minimal bound is used for the input
  distributions.}
  \label{asy_sc5}
\end{figure}

\section{Dilepton production via the $J/\psi$ resonance in the GSI regime}

To achieve a higher counting rate, one may exploit the $J/\psi$ peak, where the
cross-section is two orders of magnitude larger. If $J/\psi$ production is
dominated by $q\bar{q}$ annihilation channel, the corresponding asymmetry has
the same structure as in the continuum region, since the $J/\psi$ is a vector
particle and the $q\bar{q}-J/\psi$ couplings are similar to
$q\bar{q}-\gamma^{*}$.\cite{Anselmino1} SPS data\,\cite{corden} show the
$p\bar{p}$ cross-section for $J/\psi$ production at $s=80$\,GeV$^2$ to be about
10 times larger than the corresponding $pp$ cross-section, indicating the
dominance of the $q\bar{q}$ annihilation mechanism. Thus, the helicity structure
of the asymmetries is preserved and, replacing the couplings in
Eq.~(\ref{LOAtt}), we can write
\begin{align}
  & A_{TT}^{J/\psi} = \hat{a}_{TT}
\\
  & \quad \null \times
  \frac{\sum_{q}(g^{V}_q)^2 \left[\Delta_T q(x_1,M^2)\Delta_T q(x_2,M^2)+
  \Delta_T\bar{q}(x_1,M^2)\Delta_T\bar{q}(x_2,M^2)\right]}{\sum_{q}(g^{V}_q)^2
  \left[q(x_1,M^2)q(x_2,M^2)
  +\bar{q}(x_1,M^2)\bar{q}(x_2,M^2)\right]}
  \nonumber
\end{align}
In the large $x_1, x_2$ region the $u$ and $d$ valence quarks dominate and,
since the $q\bar{q}-J/\psi$ coupling is the same for $u$ and $d$ quarks, the
asymmetry becomes
\begin{equation}
  A_{TT}^{J/\psi} \simeq
  \hat{a}_{TT}
  \frac{\Delta_T u(x_1,M^2)\Delta_T u(x_2,M^2)
       +\Delta_T d(x_1,M^2)\Delta_T d(x_2,M^2)}
  {u(x_1,M^2)u(x_2,M^2)+d(x_1,M^2)d(x_2,M^2)}
  \,.
\end{equation}
The condition $\Delta_T u(x)\gg\Delta_T d(x)$, satisfied by all models at large
$x$, permits a further simplification and one obtains
\begin{equation}
  A_{TT}^{J/\psi} \simeq \hat{a}_{TT}
  \frac{\Delta_T u(x_1,M^2)\Delta_T u(x_2,M^2)}{u(x_1,M^2)u(x_2,M^2)}
  \,.
\end{equation}
The $J/\psi$ asymmetry is then essentially the DY asymmetry evaluated at
$M_{J/\psi}$ and, for $s = 80$\,GeV$^2$, lies in the range 0.25--0.45 (see
Fig.~\ref{asy_sc5}). Inasmuch as the $gg$ fusion diagram may be neglected, as
old $p\bar{p}$ data suggest, this remains true at NLO (i.e.\ considering gluon
radiation).

\section{Threshold resummation}

The kinematic region corresponding to $M \approx 1-4$\,GeV and with a
centre-of-mass energy $s\approx 30$\,GeV$^2$ is not properly contained in the
domain of perturbative QCD, (i.e.\ factorization, parton model etc.). Thus,
depending on kinematics, higher-order corrections to the cross-sections may be
important and must be well understood.

For the sake of simplicity, we shall merely sketch what occurs, with little
quantitative detail. The factorization theorem for the hadronic cross-section in
terms of twist-2 parton densities is not exact, but holds only to the leading
power of $M$, and the corrections generally increase as $\tau$ increases. In the
region $z=\tau/(x_1x_2)\simeq1$ the kinematics is such that virtual and
real-emission diagrams become strongly unbalanced (real-gluon emission is
suppressed) and in these conditions there are large higher-order logarithmic
corrections to the partonic cross-section of the form
\begin{equation}
  \alpha_{s}^{k}\frac{[\ln(1-z)]^{2k-1}}{(1-z)}
\end{equation}
The region $z\approx1$ is dominant in the kinematic regime relevant for GSI,
hence the large logarithmic contributions need to be resummed to all orders in
$\alpha_s$. NLL-resummed perturbation theory has been extensively
studied\,\cite{Vogelsang1} and resummation corrections for $A_{TT}$ are found to
be less than $10\%$ and rather dependent on the infrared cut-off for the soft
gluon emission.

\section{Conclusions}

In the GSI regime Drell--Yan double transverse-spin asymmetries are sizable,
of the order of $30\%$, and are not spoiled by NLO (and resummation) effects.
Transverse asymmetries for $J/\psi$ production at moderate energies are expected
to be similar (with the advantage of much higher counting rates). Transversely
polarized antiproton experiments at GSI will thus provide an excellent window
onto nucleon transversity.

\section*{Acknowledgments}
We would like to thank M.~Anselmino, N.N.~Nikolaev and our colleagues of the PAX
collaboration for prompting the study reported here and for various useful
discussions. This work is supported in part by the Italian Ministry of
Education, University and Research (PRIN~2003).

\end{document}